\documentclass[aip,jcp,reprint,groupedaddress,twocolumn]{revtex4-1}
\usepackage{graphicx}
\usepackage{epsfig}
\usepackage{amsmath}

\begin{document}

\title{Kramers problem for nonequilibrium current-induced chemical reactions }

\author{Alan A. Dzhioev}
\altaffiliation{On leave of absence from Bogoliubov Laboratory of Theoretical Physics, Joint Institute for Nuclear Research,  RU-141980 Dubna, Russia }
\author{D. S. Kosov}
\email{dkosov@ulb.ac.be}
\affiliation{Department of Physics,
Universit\'e Libre de Bruxelles, Campus Plaine, CP 231, Blvd du Triomphe, B-1050 Brussels, Belgium }


\begin{abstract}
We discuss the use of tunneling electron current to control and catalyze chemical reactions.
Assuming  the separation of time scales for electronic and nuclear dynamics we employ  Langevin equation for a reaction coordinate.
The Langevin equation contains nonconservative current-induced forces and gives nonequilibrium, effective potential energy surface for current-carrying molecular systems. The current-induced forces are computed via Keldysh nonequilibrium Green's functions.
Once a nonequilibrium, current-depended potential energy surface is defined, the chemical reaction is modeled as an escape of a Brownian particle from the potential well. We demonstrate that the barrier between the reactant and the product states can be controlled by the bias voltage.
When the molecule is asymmetrically coupled to the electrodes, the reaction can be catalyzed or stopped depending on the polarity of the
tunneling current.
\end{abstract}

\maketitle


When a  molecule is attached to two metal electrodes with different  chemical potentials  or when it is placed on a surface
under  scanning tunneling microscope (STM) tip, the electron current flows through it. It brings the molecule out of
equilibrium and changes its electronic, vibrational  and mechanical properties.
The interaction of nonequilibrium current-carrying electrons
with nuclear degrees of freedom may catalyze certain
chemical reactions which are not possible under equilibrium conditions.
This opens a possibility to use the tunneling molecular junction as  a nanoscale chemical reactor rather than an electronic element of a  circuit.
The recent experimental work has demonstrated that the chemical bonds can be selectively broken
or formed by the tunneling electron current.\cite{PhysRevLett.78.4410,PhysRevLett.84.1527,ho99}
The tunneling current can even initiate  chemical reactions in the reactants which are brought close to each other under the STM tip.\cite{PhysRevLett.85.2777,Repp26052006}
To predict the outcome of  current-induced chemical reactions and to guide the experimental work, we need to develop  intuitively simple reaction rate theory  which takes into account the tunneling molecular junction conditions.

Chemical reactions in a complex environment are traditionally modeled as an escape of a Brownian particle from a potential
well. One usually begins with the (generalized) Langevin equation for the reaction coordinate
and then  computes the rate at which an effective Brownian particle escapes from the potential well over a potential barrier
(so-called Kramers problem).\cite{zwanzig-book,nitzan-book} However, when the molecule is   driven out of equilibrium
by the tunneling flow of electrons through it,  the energy
or  free energy  surface can not be defined. How can one formulate and solve Kramers problem in this case?
That is one of the key questions which we address in this paper.


Let us consider a molecule attached to two metal electrodes. One electrode can be, for example, the metal surface and the other one is a STM tip. The molecule is modeled by one electronic spin-degenerate molecular orbital
with energy $\varepsilon(x)$, which depends on some  reaction coordinate  $x$ and the gate voltage.
The reaction coordinate   is considered to be a classical variable with corresponding momentum $p$
and reduced mass $m$. The nuclear Coulomb repulsion energy  is  $U(x)$.
Then the  molecular Hamiltonian (we use atomic units throughout the paper) is
\begin{equation}
H_M= \varepsilon(x) \sum_\sigma a^{\dagger}_\sigma a_\sigma
+ \frac{p^2}{2m}  + U(x).
\label{ham1}
\end{equation}
Here
$a^{\dagger}_\sigma$($a_\sigma $)  creates (annihilates) an electron with the spin~$\sigma$ in the molecule. The total molecular junction Hamiltonian consists of the molecular Hamiltonian (\ref{ham1}), the Hamiltonians for noninteracting left  and right electrodes, and the molecule-electrode
interaction:
\begin{equation}
H=H_M +  \sum_{\sigma, k \in L,R}  \varepsilon_k a^{\dagger}_{\sigma k} a_{\sigma k} +
\sum_{\sigma, k \in L,R } (t_k a^{\dagger}_{\sigma k} a_\sigma +\mbox{h.c.} ),
\label{ham2}
\end{equation}
where $a^{\dagger}_{\sigma k}$($a_{\sigma k}$) creates (annihilates) an electron in the state $\sigma k$ of either the left ($L$) or  the right ($R$) electrodes.
Electron creation and annihilation operators satisfy standard fermionic anticommutation relations.
Tunneling coupling matrix element  is  $t_k$.  Since the  screening length in the metallic electrodes is very short, we assume that the voltage bias drops on the interface. Therefore  there is no  external electric field in the molecule.

To derive Langevin equation for the reaction coordinate we partition the molecular junction into two parts: a "system" and a "bath". The system is the reaction coordinate $x$ and the bath consists of all electronic degrees of freedom in the molecule and electrodes.
The bath degrees of freedom can be projected out from the equations of motion and affect the reaction coordinate only through effective forces.\cite{tully95}
Furthermore,
we assume that the  time-scales for electronic and nuclear motions can be separated: The electronic degrees of freedom are much faster than  the motion of the molecule along the reaction coordinate $x$. Therefore, we can assume that the electronic steady state  is instantaneously established along the reaction coordinate trajectory $x=x(t)$. So the nonequilibrium electronic density matrix  $\rho(x)$ depends on  time only through the parametrical dependence on $x$. The result is
the Langevin equation for the reaction coordinate\cite{tully95,PhysRevLett.97.046603,fuse}
\begin{equation}\label{eq_of_mot}
m \ddot x = - \mathrm{Tr}\bigl[\rho(x) \frac{\partial H}{\partial x}\bigr]  -  \zeta \dot x + \delta f(t).
\end{equation}
Here the conservative part of the force is given by  the nonequilibrium analogue of the Hellmann-Feynman theorem $\mathrm{Tr}[\rho(x) \frac{\partial H}{\partial x} ]$,\cite{diventra:16207}
and  $\delta f(t)$  (fluctuating force),   $\zeta \dot x$  (frictional force) are nonconservative  contribution originated from the integrated out electronic degrees of freedom.\cite{tully95}
The nonconservative  forces describe  Joule heating, i.e. the energy loses due to particle-hole excitations in the molecule and metal electrodes.
For particular systems, the conservative current-induced forces can be computed with the use of  nonequilibrium Green's functions within
tight-binding approximation or density functional theory \cite{todorov01,PhysRevB.67.193104}.

The noise is taken in the Gaussian form and  related to the viscosity by the fluctuation-dissipation relation with some temperature $T$:
\begin{equation}
\langle \delta f(t) \rangle =0,~~~~ \langle \delta f(t) \delta f(t') \rangle =2 \zeta T \delta(t-t').
\label{fldis}
\end{equation}
We assume  that the electrons and vibrations have the same temperature $T$
although we fully appreciate that far from equilibrium  vibrational temperature can deviate from electronic
temperature \cite{PhysRevB.69.245302,thoss11,PhysRevB.73.155306}.
Since  $ \partial H/ \partial x  = \varepsilon'(x)\sum_\sigma a^{\dagger}_\sigma a_\sigma + U'(x)$,  Eq.~\eqref{eq_of_mot}  becomes
\begin{equation}
m \ddot x = - \varepsilon'(x) n(x)-  U'(x)  - \zeta \dot x + \delta f(t) - \zeta \dot x,
\label{langevin}
\end{equation}
where $n(x) = 2\mathrm{Tr}[\rho(x)  a^{\dagger}_\sigma a_\sigma ] $ is the nonequilibrium  population of the electronic level in the molecule.

To complete Langevin equation for the reaction coordinate (\ref{langevin}) we need to know the explicit expression for $n(x)$.
For a given value of $x$  the population $n(x)$ can be computed by Keldysh nonequilibrium Green's functions.\cite{keldysh65}
The derivations are relatively straightforward, so we just outline them here without giving the full details.\cite{haug-jauho}
One begins with Keldysh contour-ordered Green's function and writes the Dyson equation for it. Applying the Langreth rules for analytical continuations
the Dyson equation is solved for and the nonequilibrium population and electron current are associated with $G^{<}$ (lesser) Green's function on the real time axis.
It results in the following expression:\cite{haug-jauho}
\begin{equation}
\label{n}
  n(x)= 2 \int \frac{d\omega}{\pi} \frac{\Gamma_L(\omega) f_L(\omega)+ \Gamma_L(\omega) f_R(\omega)}{(\omega-\varepsilon(x)  -\Lambda (\omega))^2 +(\Gamma(\omega))^2}.
\end{equation}
Here $\Gamma=\Gamma_L+\Gamma_R$ and  $\Lambda=\Lambda_{L}+\Lambda_{R}$ determine the broadening and shift of the molecular level due to coupling to the electrodes. They are given
by the real and imaginary parts of the electrode self-energy
\begin{equation}
\label{SelfEnergy}
\Sigma_{L,R} =  \sum_{k\in L,R}\frac{|t_k|^2}{\omega-\varepsilon_{k}+i 0}= \Lambda_{L,R}  - i\Gamma_{L,R}.
 \end{equation}
The function $f_{L,R}(\omega) = [1 + e^{(\omega -\mu_{L,R})/T}]^{-1}$ is the Fermi-Dirac electron distribution in the left and right electrodes.

The corresponding electron current also depends parametrically on the reaction coordinate and can be readily calculated by the Landauer formula.\cite{haug-jauho}
In the equilibrium, i.e. when the chemical potential of the left electrode equals to the chemical potential of the right electrode, the current is zero and
Langevin equation  (\ref{langevin}) remains the same but $n(x)$ becomes the equilibrium electronic population $n_{\mathrm{eq}}(x)$, which can be computed from (\ref{n}) by setting $f_L=f_R$.

Let us introduce  the time-dependent probability distribution for the reaction coordinate  $F=F(x,p,t)$.
The Langevin equation (\ref{langevin}) is equivalent to the standard Fokker-Planck equation for the  distribution function:
\begin{equation}
\label{fk}
\frac{\partial}{\partial t}  F = - \frac{p}{m} \frac{\partial}{\partial x} F + \frac{\partial}{\partial p} (U{\,}'_{eff}(x) + \zeta \frac{p}{m} ) F + \zeta T \frac{\partial^2}{\partial p^2} F.
 \end{equation}
Here the effective nonequilibrium potential energy surface is defined via the integration of the
 nonequilibrium force in  Langevin equation~(\ref{langevin})
 \begin{equation}\label{Ueff}
 U_{eff}(x) =U(x) +   \int_{x_0}^{x} dy \; \varepsilon'(y) n(y).
 \end{equation}
The choice of $x_0$ is not relevant, since the  Fokker-Planck  equation does not depend on it.
To elucidate effects related to the electron current we separate  the effective potential into two parts:
\begin{equation} \label{U_eff_noneq}
 U_{eff}(x) =U_{\mathrm{eq}}(x) +  \int_{x_0}^{x} dy \; \varepsilon'(y) \Delta n (y).
 \end{equation}
Here the first term describes the equilibrium potential energy surface and the second term gives the nonequilibrium corrections.
The equilibrium potential energy surface, $U_{\mathrm{eq}}(x)$, is the potential energy of the molecule in the absence of electron current and it includes
equilibrium charge transfer between the molecule and metal electrodes.
Since the derivative $\varepsilon'(y)$ does not depend  on the applied voltage and on the coupling to the electrodes, the nonequilibrium correction
is  only due to the variation of the molecule population $ \Delta n(x)= n(x) - n_{eq}(x)$ caused by the current flow.
As one sees from Eq.(\ref{n}) the nonequilibrium correction $\Delta n(x)$ is most significant
when the molecular level is in resonance with the Fermi energy of the electrodes, i.e., $\varepsilon(x)\approx\mu_{L,R}$. In this case the current reaches its maximal value.
Moreover, at asymmetric molecule-electrode coupling, $\Gamma_L\ne\Gamma_R$,
$\Delta n(x)$ can be both positive and negative, depending on the polarity of the applied voltage bias. Therefore, the current flows through the molecule can locally increase or decrease
the nonequilibrium potential energy surface.

\begin{figure*}[t!]
 \begin{centering}
\includegraphics[width=16cm]{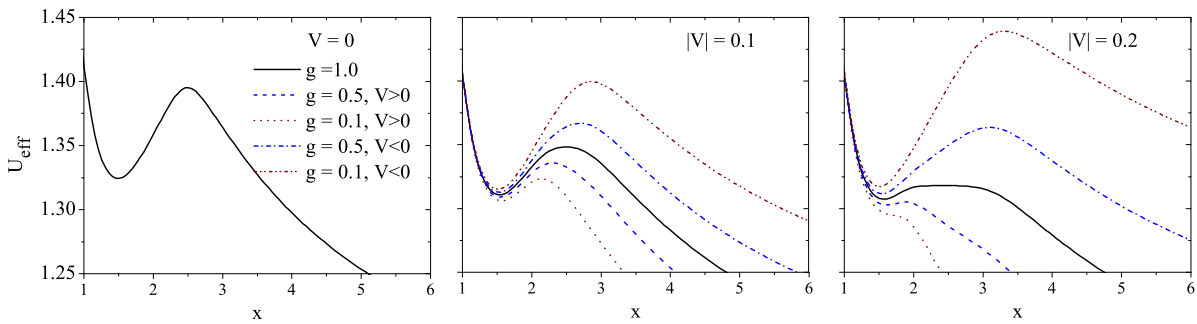}
\caption{Effective nonequilibrium potentials $U_{eff}(x)$ as functions of the reaction coordinate.
The voltage bias is $V$ and the asymmetry coefficient $g=\Gamma_L/\Gamma_R$.}
\label{fig1} \end{centering}
\end{figure*}


Let us understand the behavior of the effective potential energy surface based on numerical calculations.
We take $\varepsilon(x)$ in the form of bonding orbital for $H_2^+$ molecule~\cite{mcquarrie-qc}.
For simplicity, we assume that  both electrodes have the constant density of states, $\eta_R$ and $\eta_L$, and they are characterized by the same half-bandwidth  $D=50$.
We also assume  that the tunneling coupling $t_k$ is real and independent of~$k$.
Then  the real and imaginary parts of the electrodes self-energy are
\begin{equation}
\Lambda_{L,R}(\omega)=\frac{\Gamma_{L,R}}{\pi}\ln\left|\frac{\omega+D}{\omega-D}\right|,~~~ \Gamma_{L,R}(\omega) = \Gamma_{L,R} \Theta(D - |\omega|),
\end{equation}
where $\Gamma_{L,R}=\pi t_{L,R}^2 \eta_{L,R}$.
The total width of the molecular level $\Gamma = \Gamma_L+\Gamma_R=0.05$ is fixed in our calculations but we vary relative contributions of left and right electrodes.
The ratio $g = \Gamma_L/\Gamma_R$ is the asymmetry coefficient.
The asymmetry coefficient can be experimentally controlled by changing the relative strength of the coupling $t_k$ (\ref{ham2}) of the molecule to left/right electrodes. This for example can be accomplished by selecting chemically different left and right molecular-electrode linkers  \cite{li:9893, LiZ._jp065120t} or simply by changing the distance between the molecule and one of the electrodes. The latter can be easily accomplished on STM experiments.

Fermi energies of both electrodes coincide with the  band centers, $\varepsilon_f=0$, and
the chemical potentials are  functions of the external applied voltage  $\mu_{L,R}=\pm 0.5V$.
 The calculations are performed at the room temperature $T=300$K.
 The value of the gate voltage is chosen in  such a  way that
$\varepsilon(x_\mathrm{min})=\varepsilon_f$, where $x_\mathrm{min}=2.493$ determines  the minimum of   $H_2^+$  ground state energy.~\cite{mcquarrie-qc}
So the molecular level is in resonance with the Fermi energy of the electrodes.
This particular choice of the gate voltage is  not critical, all  our results will qualitatively remains the same for other values of
the gate voltage.
When $x<x_\mathrm{min}$ the molecular orbital
is below the Fermi energy of the electrodes, and  when $x>x_\mathrm{min}$ it lays  above it.

Fig.~\ref{fig1} shows effective nonequilibrium potential $U_{eff}$ as a function of  reaction coordinate $x$ for different values
of the applied voltage $V$ and the asymmetry coefficient $g$.
As one can see from Fig.~\ref{fig1}, the height of the potential barrier, $\Delta U_{eff} = U_{eff}(b) - U_{eff}(a)$, between product and reactant states can be made smaller or larger as the current flows through the molecule. For the symmetric coupling to the left and right electrodes, the barrier is always decreased by the voltage bias. This effect is mainly caused by the reduction of the electronic population.
The asymmetric case is much more interesting.
Here, the effect of the barrier reduction can be amplified or reversed depending on the polarity of the applied voltage bias.
We consider the case when the molecule is coupled stronger to the right electrode ($g<1$).
When the applied voltage is positive $V>0$, i.e., the left chemical potential is larger than the right chemical potential,
molecular electrons are  more depleted than in the symmetric case. Moreover, as follows from Fig.~\ref{fig1}, the applied
voltage shortens the distance $\Delta x=b-a$ in reaction coordinate  between the minimum and the top of the barrier of the potential energy surface. Both effects collectively reduce the potential barrier height $\Delta U_{eff}$ stronger than in a case of the symmetric coupling to the electrodes.
For large voltage $V=0.2$ and notable asymmetry $g=0.1$ the barrier completely vanishes.
It is interesting that by  reversing the voltage bias we can increase the barrier between the product and reactant states. For example, as one can see from Fig.~\ref{fig1}, when the left electrode is negatively biased ($V=-0.2$) and weakly coupled to the molecule ($g=0.1$) the barrier is increased by 70\% as compared to the equilibrium.
We emphasize that there is no electric field across the molecule from the voltage bias in our model, therefore the observed physical behavior is solely due to tunneling  current and corresponding nonequilibrium changes in molecular electronic population.

Let us now compute the rates for current induced chemical reactions. Since we have already computed nonequilibrium effective potential energy surface, we can use standard reaction rate theory for our calculations.
We consider separately two cases: overdamped ( $\zeta \gg \zeta_0$, where $ \zeta_0 = 2 \sqrt{m U_{\mathrm{eq}}^{''}(a)}$ ) and underdamped ($\zeta \ll \zeta_0$).
In the overdamped limit, Fokker-Plank equation (\ref{fk}) becomes   one-dimensional Smoluchowski equation for the probability distribution and the reaction rate
 $k$ to overcome the barrier can be computed by numerical integration\cite{zwanzig-book}. For small $T$ the reaction rate
can be evaluated analytically by performing the quadratic expansion of the effective potential $U_{eff}$ near the minimum and  the maximum of the barrier.\cite{zwanzig-book} It results into the standard expression for the reaction rates
\begin{equation}\label{rate}
k = \frac{\sqrt{-U''_{eff}(a) U''_{eff}(b)}}{2 \pi \zeta}  e^{-\Delta U_{eff}/T}.
\end{equation}
In the underdamped limit, the reaction rate can be also computed analytically \cite{zwanzig-book} and it is given by the following expression
\begin{equation}\label{rate2}
{k} = \frac{\zeta  \Delta U_{eff}}{ m T} e^{-\Delta U_{eff}/T}.
\end{equation}
\begin{figure}[t!]
 \begin{centering}
\includegraphics[width=\columnwidth]{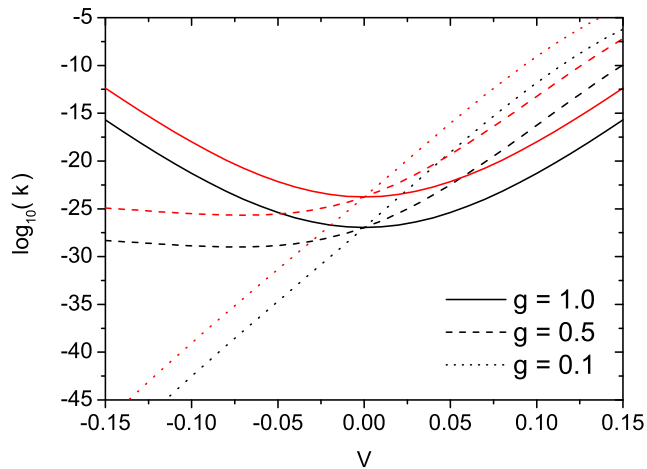}
\caption{Nonequilibrium reaction rates computed for various values of the asymmetry coefficient $g$ as a function of the applied voltage. Reaction rates are computed in underdamped (red curves) $\zeta = 0.1\zeta_0$ and overdamped (black curves) $\zeta = 10\zeta_0$  regimes.}
\label{fig2}
 \end{centering}
\end{figure}
Using Eqs.~(\ref{rate},\ref{rate2})  we calculate the reaction rate~$k$ as a function of the applied voltage bias for different values of $g\le1$. The results are shown in Fig.~\ref{fig2}.
For symmetric case ($g=1$) the reaction rate does not depend on the polarity of voltage and it increases with applied voltage.
The situation changes dramatically when the molecule is asymmetrically coupled to the electrodes.   The reaction is "catalyzed" if the voltage is positive
and  slows down if we reverse the direction of the voltage. It is interesting that the voltage dependence of the reaction rates are almost identical for overdamped and underdamped cases.  This can be easily understood  from the following geometrical consideration. Depending on the polarity, the tunneling current makes the potential barrier $\Delta U_{eff}$ larger or smaller. If the potential barrier  increases or decreases, the frequencies  $U''_{eff}(a)$ and $-U''_{eff}(b)$ increase or decrease, respectively, too. It leaves the ratio between overdamped and underdamped reaction rates almost voltage independent.


 In conclusion, we have demonstrated that the electric current which flows through the molecule can be used to control chemical reactions.
We combined Langevin equation for a reaction coordinate with Keldysh Green's function calculations of the current induced forces and demonstrated how
the nonequilibrium, current-depended potential energy surface can be defined. The chemical reaction is modeled as an escape of a Brownian particle from the potential well (Kramers problem). The barrier between the reactant and the product states can be controlled by the bias voltage.
We demonstrated that
when the molecule is asymmetrically coupled to the electrodes the reaction can be catalyzed or stopped depending upon the direction of the electric current which flows through the molecule.

We thank Maxim Gelin for many valuable discussions.
 This work has been supported by the Francqui Foundation, Belgian Federal Government under the Inter-university Attraction Pole project NOSY  and
 Programme d'Actions de Recherche Concert\'ee de la Communaut\'e francaise (Belgium) under project "Theoretical and experimental approaches to surface reactions".

\end{document}